\documentstyle[preprint,aps]{revtex}
\begin{document}
\draft
\title{On Thermometer Operation of Ultrasmall Tunnel Junctions}
\author{Heinz-Olaf Mller, Andreas H"dicke, and Wolfram Krech}
\address{Institut fr Festk"rperphysik\\
        Friedrich--Schiller--Universit"t Jena\\
        Max--Wien--Platz 1\\
        D-07743 Jena}
\date{\today}
\maketitle
\begin{abstract}
The temperature dependence of the $I$--$V$ characteristics of many
single--electron tunneling devices enable thermometer operation
of these systems. We investigate two normal conducting kinds of
them, {\sl (a)} a single junction in a high--impedance environment,
and {\sl (b)} a double junction. The characteristics of both devices
show a crossover from Coulomb blockade at low temperatures to ohmic
behavior at high temperatures. The related differential conductivity
dip allows the determination of the junctions temperature. Both
configurations {\sl (a)} and {\sl (b)} are expected to work at
least within the range $0.5\le\beta E_{\rm C}\le 10$, where $E_{\rm C}$
is the Coulomb energy of the system under investigation. We present an
analytical solution for both low-- and high--temperature case of
{\sl (a)} and {\sl (b)} as well as numerical results and their fit,
including the effect of co--tunneling in case of a double junction.
\end{abstract}

\section{Introduction}
\label{intro}

Sensoring turns out to be one of the possible applications of
single--electron tunneling (SET) devices. After the establishment of the
electrometer~\cite{kor1}, that is one of the basic devices of SET,
radiation detectors~\cite{ave4} and resistor comparators~\cite{hom1} were
studied. Nowadays, SET--devices as thermometer are brought into the scope
of investigation~\cite{pek1}.

For this operation the known temperature dependence of the current
through ultrasmall tunnel junctions is exploited. As the single junction
without any environment, i.e.\ the voltage biased tunnel junction, does
not show a reasonable influence of the temperature on its characteristic,
the basic thermometer devices are {\sl (a)} the normal conducting
single junction in a high--impedance environment and {\sl (b)} the
normal conducting double junction. Studies on both {\sl
(a)}~\cite{ing3} and {\sl (b)}~\cite{pek1} are done. This paper is
restricted to {\sl (a)} the infinite high--impedant ohmic
environment ($R_{\rm E}\to\infty$) and {\sl (b)} the symmetric
double junction ($R_1=R_2$, $C_1=C_2$). On the other hand, it
extends~\cite{ing3} because of spotting the thermometer operation as
well as~\cite{pek1} due to a more rigid theoretical treatment.

The suppression of thermal fluctuations in SET junctions is expressed by
the condition $E_{\rm C}\gg k_{\rm B}T$, where $E_{\rm C}$ is the Coulomb
energy of the considered system. If this condition holds both {\sl (a)}
and {\sl (b)} show a Coulomb blockade~\cite{dev1},~\cite{ave1}. This is
the low--temperature case of our consideration. At high temperatures,
$E_{\rm C}\ll k_{\rm B}T$, Coulomb effects are covered by thermal
fluctuations and an ohmic current--voltage characteristic remains.

The thermal dependence is rather discussed in terms of the
differential conductivity $G(V)$ of the junction(s) than the current.
According to the Coulomb blockade $G(V)$ shows a dip at $V=0$
(Fig.~\ref{fig1},~\ref{fig3}). This dip is well--pronounced at low
temperatures and is smeared out with increasing temperatures. Therefore,
the shape of the dip allows the determination of the temperature of the
junctions. This is the thermometer operation.

There are different measures of the mentioned dip that may be used for
evaluation. In this paper the depth of the dip, i.e.\ $G(V=0)$, is
used only. It is preferred to the width of the dip because a precise
measurement of the latter at higher temperatures would require
improper high voltages across the junction(s). Additionally, the depth of
the dip is easier accessible in theory.

\section{Single Junction}
\label{single}

A single junction possesses the Coulomb energy $E_{\rm C} = e^2/(2C)$,
where $C$ is the junction capacitance. The current via this junction in
a high--impedance ohmic environment ($R_{\rm E}\to\infty$) is given
by~\cite{dev1}
\begin{equation}\label{sint}
I(V) = \frac{1}{e\,R}\Big(1-{\rm e}^{-\beta e\,V}\Big)
\sqrt{\frac{\beta}{4\pi E_{\rm C}}}\int{\rm d}\varepsilon
\,\frac{\varepsilon}{1-{\rm e}^{-\beta\varepsilon}}\,
\exp\Big[-\frac{\beta}{4 E_{\rm C}}\Big(e\,V-E_{\rm C}-\varepsilon
\Big)^2\Big]
\end{equation}
in terms of the junction resistance $R$ and the temperature $\beta =
1/(k_{\rm B}T)$. To calculate this integral the first term of it is
represented as
\begin{equation}
\frac{\varepsilon}{1-{\rm e}^{-\beta\varepsilon}} = \left\{
\begin{array}{c@{\quad{\rm if}\quad}l}
\varepsilon\sum\limits_{n=0}^{\infty}{\rm e}^{-n\beta\varepsilon} &
\varepsilon>0\\
-\varepsilon\sum\limits_{n=0}^{\infty}{\rm e}^{(n+1)\beta\varepsilon} &
\varepsilon<0.
\end{array} \right.
\end{equation}
This representation breaks down at $\beta=0$. Furthermore, the
convergence of the sum is poor in the vicinity of this value.
Therefore, this domain is excluded now but studied in detail later.
Due to the sum, (\ref{sint}) simplifies to two integrals of the type
\begin{equation}
\int\limits_{0}^{\infty}{\rm d}x\,x\,{\rm e}^{-(a\,x^2+2b\,x+c)}
= \frac{1}{2a}\,{\rm e}^{-c}
-\frac{b}{2a}\sqrt{\frac{\pi}{a}}\exp\Big(\frac{b^2}{a}-c\Big)
{\rm erfc}\Big(\frac{b}{\sqrt{a}}\Big),
\end{equation}
with ${\rm erfc}(x) = 1 - {\rm erf}(x)$. The corresponding calculation
yields
\begin{equation}
I(V) = \frac{E_{\rm C}}{2e\,R}\,\sum\limits_{n=0}^{\infty}
\Big[{\rm f}\Big(\frac{e\,V}{E_{\rm C}},n\Big)-{\rm f}\Big(-\frac{e\,V}
{E_{\rm C}},n\Big)\Big]
\end{equation}
with
\begin{eqnarray}
{\rm f}(x,n) & = & {\rm e}^{-\beta E_{\rm C}(x-1)^2/4}\Bigg[
\frac{4}{\sqrt{\pi\beta E_{\rm C}}}\\
& & -(2n-x+1)\exp[\beta E_{\rm C}(2n-x+1)^2/4]
{\rm erfc}\Big((2n-x+1)\sqrt{\frac{\beta E_{\rm C}}{4}}\Big)\nonumber\\
& & -(2n+x+1)\exp[\beta E_{\rm C}(2n+x+1)^2/4]
{\rm erfc}\Big((2n+x+1)\sqrt{\frac{\beta E_{\rm C}}{4}}\Big)
\Bigg].\nonumber
\end{eqnarray}
Obviously, the current is the difference of two contributions depending
on each other by the substitution $V\leftrightarrow -V$. The
determination of the minimal differential conductivity yields in terms
of
\begin{eqnarray}
\frac{{\rm d}}{{\rm d} y}
\left.\Big[{\rm f}(y,n)-{\rm f}(-y,n)\Big]\right|_{y=0}
& = & 4\sqrt{\frac{\beta E_{\rm C}}{\pi}}\,{\rm e}^{-\beta E_{\rm C}/4}\\
& &
-2\beta E_{\rm C}(2n+1)\,{\rm e}^{\beta E_{\rm C}n(n+1)}
{\rm erfc}\Big((2n+1)\sqrt{\frac{\beta E_{\rm C}}{4}}\Big)
\nonumber
\end{eqnarray}
for $G(V)$
\begin{eqnarray}\label{gv0}
G(V=0) & = & \left.\frac{{\rm d}I}{{\rm d}V}\right|_{V=0}\\
& = & \frac{2}{R}\,{\rm e}^{-\beta E_{\rm C}/4}\sum_{n=0}^{\infty}
\left[\sqrt{\frac{\beta E_{\rm C}}{\pi}}-\beta E_{\rm C}(n+\frac{1}{2})
\,{\rm e}^{\beta E_{\rm C}(n+1/2)^2}
{\rm erfc}\Big((n+\frac{1}{2})\sqrt{\beta E_{\rm C}}\Big)\right].
\nonumber
\end{eqnarray}
If $\beta E_{\rm C}>1$ holds the asymptotic expansion~\cite{abr1}
\begin{equation}
\sqrt{\pi}z\,{\rm e}^{z^2}{\rm erfc}(z) = 1+\sum_{m=1}^{\infty}
(-1)^m\frac{1\cdot3\cdot\ldots\cdot(2m-1)}{(2z^2)^m}
\end{equation}
and~\cite{gra4}
\begin{equation}
\sum_{n=0}^{\infty}(2n+1)^{-2m} = \frac{\left(2^{2m}-1\right)\pi^{2m}}
{2(2m)!}\left|{\rm B}_{2m}\right|
\end{equation}
can be used to transform (\ref{gv0}) into
\begin{equation}\label{gv01}
G(V=0) = -\frac{1}{R}\sqrt{\frac{\beta E_{\rm C}}{\pi}}\,
{\rm e}^{-\beta E_{\rm C}/4}\sum_{m=1}^{\infty}(-1)^m
\frac{\left(2^{2m}-1\right)\pi^{2m}}{\left(\beta E_{\rm C}\right)^m m!}
\left|{\rm B}_{2m}\right|
\end{equation}
using the Bernoulli numbers ${\rm B}_{2m}$. The term corresponding to
$m=1$ yields the behavior of (\ref{gv01}) for $\beta E_{\rm C}\gg1$.
This is the short--dashed curve in Fig.~\ref{fig2}.

In case of $\beta E_{\rm C}\ll 1$ the evaluation of (\ref{sint}) uses
\begin{equation}
\frac{\varepsilon}{1-{\rm e}^{-\beta\varepsilon}} =
\frac{1}{\beta}\,{\rm e}^{\beta\varepsilon/2}\frac{\beta\varepsilon/2}
{\sinh\beta\varepsilon/2}.
\end{equation}
Therefore, the integral in (\ref{sint}) includes besides the Gaussian an
other peak of the shape $x/\sinh x$. Because the width of the Gaussian
and the $x/\sinh x$--peak are proportional to $\left(\beta E_{\rm
C}\right)^{-1/2}$ and $\left(\beta E_{\rm C}\right)^{-1}$, respectively.
Hence, the latter peak becomes broad and flat compared to the Gaussian in
the limit $\beta\to0$. Therefore, it turns out to be sufficient to use
the approximation
\begin{equation}
\frac{\beta\varepsilon/2}{\sinh\beta\varepsilon/2}\approx 1
\end{equation}
in this limit and an integral of the Gaussian remains of (\ref{sint}).
It results in
\begin{equation}
I(V) = \frac{1}{\beta e\,R}\Big(1-{\rm e}^{-\beta e\,V}\Big)
\exp\Big[-\frac{\beta}{2}\big(E_{\rm C}-2e\,V\big)\Big]
\end{equation}
for the current via the junction and gives
\begin{equation}
G(V=0) = \left.\frac{{\rm d}I}{{\rm d}V}\right|_{V=0}
= \frac{1}{R}\,{\rm e}^{-\beta E_{\rm C}/4}
\end{equation}
for the minimal differential conductivity of the junction. This formula is
represented
in Fig.~\ref{fig2} by means of a long--dashed curve.

According to Fig.~\ref{fig2}, an evaluation of (\ref{sint}) at $\beta
E_{\rm C}\approx 1$ is desired, but hardly to achieve analytically.
Therefore, a simple fit to the numerical data was performed, resulting in
\begin{equation}\label{fit1}
G(V=0) = \frac{1}{R}\Big(1-\frac{2}{\pi}\arctan(0.5\beta E_{\rm C})\Big).
\end{equation}
This formula corresponds to the dotted line in Fig.~\ref{fig2}. The fit
may be enhanced by introducing a polynomial into the argument of
$\arctan$. In the given simple form (\ref{fit1}) enables an inversion to
\begin{equation}
\beta E_{\rm C} = \frac{1}{0.5}
\tan\Big[\frac{\pi}{2}\Big(1-G(V=0)R\Big)\Big],
\end{equation}
a simple thermometer function.

\section{Double Junction}
\label{double}

In case of the symmetric double junction the Coulomb energy is given by
$E_{\rm C} = e^2/(4C)$, where $C$ is the junction capacitance of each
junction. Due to several states of the central electrode of the double
junction, that are determined by the number $n$ of extra charges on it,
the current $I(V)$ turns out to be a sum over contributions of each
state~:
\begin{equation}\label{ssum}
I(V) = e\sum_n\Big[r_1(n)-l_1(n)\Big]\sigma(n)
= e\sum_n\Big[r_2(n)-l_2(n)\Big]\sigma(n)
= \sum_ni(n)\sigma(n).
\end{equation}
In general, both the rates $r_{1,2}(n)$, $l_{1,2}(n)$ for both directions
of the first and second junction and the occupation probability
$\sigma(n)$ are voltage dependent. The voltage dependence of the rates is
given in terms of the energy gain per tunnel event~\cite{ave3}
\begin{equation}
E^{\,r}_{1,2}(n) = 2E_{\rm C}\Big(\frac{C\,V}{e}\mp n-\frac{1}{2}\Big);
\hspace{2cm}
E^{\,l}_{1,2}(n) = 2E_{\rm C}\Big(-\frac{C\,V}{e}\pm n-\frac{1}{2}\Big)
\end{equation}
by
\begin{equation}
r_{1,2}(n) = \frac{1}{e^2R}\,\frac{E^{\,r}_{1,2}(n)}{1-{\rm e}^{-\beta
E^{\,r}_{1,2}(n)}};
\hspace{2cm}
l_{1,2}(n) = \frac{1}{e^2R}\,\frac{E^{\,l}_{1,2}(n)}{1-{\rm
e}^{-\beta E^{\,l}_{1,2}(n)}},
\end{equation}
where $R$ is the junction resistance.

According to the Likharev--Averin equation~\cite{ave3} the
stationary occupation probabilities $\sigma(n)$ are governed by
\begin{equation}
\frac{\sigma(n+1)}{\sigma(n)} = \frac{r_1(n)+l_2(n)}{r_2(n+1)+l_1(n+1)}.
\end{equation}
In case of $n\gg 1$ this simplifies to
\begin{equation}
\frac{\sigma(n+1)}{\sigma(n)}
\approx\,{\rm e}^{-2n\beta E_{\rm C}}\cosh 2\beta E_{\rm C}
\Big(\frac{C\,V}{e}-\frac{1}{2}\Big).
\end{equation}
Comparing with a Gaussian normal probability function with mean value $0$
and width $s$
\begin{eqnarray}
\sigma(n) & = & \frac{1}{s\sqrt{\pi}}\exp\Big[-\frac{1}{2}\Big(\frac{n}{s}
\Big)^2\Big]\\
\frac{\sigma(n+1)}{\sigma(n)} & = & \exp\Big[-\frac{n}{s^2}-\frac{1}{2s^2}
\Big],\nonumber
\end{eqnarray}
the occupation probability can be approximated by this Gaussian, if
\begin{equation}
s = \frac{1}{\sqrt{2\beta E_{\rm C}}}
\end{equation}
is chosen. This approximation is in the limit $\beta\to0$ exact.
However, according to Fig.~\ref{fig5} it seems to be reasonable, if
$\beta E_{\rm C} < 10$ holds. The main result of this consideration is
the weak voltage dependence of the occupation probabilities $\sigma(n)$.
Neglecting this dependence, the differential conductivity of the studied double
junction might be expressed as
\begin{equation}\label{gsum}
G(V) = \frac{{\rm d}I(V)}{{\rm d}V} = \sum_ng(n)\sigma(n),
\end{equation}
where
\begin{equation}
g(n) = \frac{{\rm d}i(n)}{{\rm d}V}
\end{equation}
is used with regard to (\ref{ssum}).
Investigating the minimal differential conductivity of the double junction,
i.e.
$V=0$, the calculation yields
\begin{eqnarray}
\left.g(n)\right|_{V=0} & = & \frac{1}{2R}\left\{1-
\frac{\sinh\beta E_{\rm C}+\beta E_{\rm C}\cosh\beta E_{\rm C}}
{\cosh\beta E_{\rm C}-\cosh2n\beta E_{\rm C}}\right.\\
& & \left.-\frac{\left[2n\beta E_{\rm C}\sinh2n\beta E_{\rm C}-\beta
E_{\rm C}\sinh\beta E_{\rm C}\right]\sinh\beta E_{\rm C}}
{\left[\cosh\beta E_{\rm C}-\cosh2n\beta E_{\rm
C}\right]^2}\right\}.\nonumber
\end{eqnarray}
In case of $n=0$ this result simplifies to
\begin{equation}\label{gn0}
\left.g(0)\right|_{V=0} = \frac{1}{R}
\frac{1-{\rm e}^{\beta E_{\rm C}}+\beta E_{\rm C}\,{\rm e}^{\beta E_{\rm
C}}}{\left({\rm e}^{\beta E_{\rm C}}-1\right)^2}.
\end{equation}
Further simplification is achieved in the case of $\beta E_{\rm C}\ll1$,
and $\beta E_{\rm C}\gg1$ where
\begin{eqnarray}\label{g0}
\beta E_{\rm C}\ll1 & : & g(0)\approx\frac{1}{2R}\left(1-
\frac{\beta E_{\rm C}}{3}\right)\\
\beta E_{\rm C}\gg1 & : & g(0)\approx\frac{\beta E_{\rm C}}{R}\,
{\rm e}^{-\beta E_{\rm C}}\nonumber
\end{eqnarray}
holds, respectively. In both cases further evaluation is possible owing
to (\ref{gsum}) and the normalization of $\sigma(n)$,
\begin{equation}
\sum_n\sigma(n) = 1.
\end{equation}
As $g(n)$ turns out to be independent of $n$ up to second order in
$\beta E_{\rm C}$ the approximation $g(n)\approx g(0)$ can be exploited
in the first case ($\beta E_{\rm C}\ll1$). This results in
\begin{equation}
G(V=0) \approx g(0)\sum_n\sigma(n) = g(0) = \frac{1}{2R}
\left(1-\frac{\beta E_{\rm C}}{3}\right).
\end{equation}
This formula is represented by the long--dashed curve in Fig.~\ref{fig4}.
In the opposite limit, i.e.\ $\beta E_{\rm C}\gg1$, $\sigma(n)$ is
sharply peaked at $n=0$ (see Fig.~\ref{fig5}$(c)$) and the evaluation
yields
\begin{equation}
G(V=0) \approx g(0)\sigma(0) = g(0) = \frac{\beta E_{\rm C}}{R}\,
{\rm e}^{-\beta E_{\rm C}}.
\end{equation}
This result is shown by the short--dashed curve in Fig.~\ref{fig4}.

Again, the missing analytical approximation at $\beta E_{\rm C}$ is
substituted by the following fit~:
\begin{eqnarray}\label{fit2}
G(V=0) & = & \frac{1}{2R}[1-\tanh(0.3\beta E_{\rm C})]\\
\beta E_{\rm C} & = & \frac{1}{0.3}\,{\rm artanh}[1-2G(V=0)R]\nonumber
\end{eqnarray}
This fit is given by the dotted curve in Fig.~\ref{fig4}. The second line
of (\ref{fit2}) enables the determination of the temperature out of the
minimal differential conductivity $G(V=0)$.

\section{Co--Tunneling}

According to~\cite{ave8} the correlated transfer of electrons across
both junctions of a double junction results in current contributions
due to two different processes known as coherent and incoherent
co--tunneling. The influence of the coherent co--tunneling can be
neglected in this consideration because it depends on the size of the
central electrode which is a design parameter. Hence, it should not be
a problem to suppress this influence by means of a larger island
between the junctions.

On the other hand, the current contribution of incoherent
co--tunneling is independent of this size. It is given by~\cite{ave8}
\begin{equation}
I_{\rm in}(V) = \frac{\hbar}{12\pi e^2R_1R_2}
\Big(\frac{1}{\Delta E_1}+\frac{1}{\Delta E_2}\Big)^2\Big((e\,V)^2
+(2\pi k_{\rm B}T)^2\Big)V.
\end{equation}
The evaluation in case of the symmetric double junction ($R_1=R_2=R$,
$\Delta E_1|_{V=0} = \Delta E_2|_{V=0}=E_{\rm C}$) at $V=0$ yields for
the contribution to the differential conductance
\begin{equation}\label{34}
G_{\rm in}(V=0) = \frac{2}{3R}\,\frac{R_{\rm Q}}{R}\,\frac{1}{\beta
E_{\rm C}}
\end{equation}
using the abbreviation $R_{\rm Q} = h/e^2$. In order to calculate the
total differential conductance at $V=0$, this term has to be added to
the results of Sec.~\ref{double}. Hence, the thermometric use
of the double junction at higher temperatures ($\beta E_{\rm C}\ll1$)
requires high tunnel resistances. Fig.~\ref{fig6} shows the
high--temperature approximation of the conductance $G(V=0)$ for
different tunnel resistances. As seen from Eq. (\ref{34})
co--tunneling becomes negligible for low temperatures ($\beta
E_{\rm C} \gg 1$). In case of a realistic value $R\approx
10R_{\rm Q}$ incoherent co--tunneling restricts the application to
$\beta E_{\rm C}\ge 0.5$. If co--tunneling reaches the order of the
one--particle process, the calculation which bases on a perturbation
theory becomes insecure.

\section{Discussion}
\label{dis}

There are obvious similarities in the behavior of the devices of
Sec.~\ref{single} and Sec.~\ref{double} to build a simple thermometer
on base of SET junctions as far as co--tunneling is excluded. In this
case the characteristics $G(\beta E_{\rm C})$ are comparable, but due
to co--tunneling that rises strongly with growing temperature the
working field of the double junction is restricted to $0.5\le\beta
E_{\rm C}\le 10$ approximately, whereas the single junction is expected
to work reasonable within $0.1\le\beta E_{\rm C}\le 10$. Therefore, the
single junction is preferred from a theoretician's point of view.

In experimental realization, however, it is much harder to prepare a
current biased single junction than a voltage biased double junction.
Hence, an experimenter would prefer the second kind of SET--thermometer.
According to these points the cons and pros of both configurations are
balanced. It could be expected that multi junction arrays are better
suited at high temperatures ($\beta E_{\rm C}\le 1$) because of
suppressed co--tunneling.

Practical exploitation of SET thermometers requires the knowledge of
the junction parameters, i.e.\ the junction capacitance and resistance.
These parameters might be determined by measuring the $I(V)$
characteristic of the system at low temperatures $\beta E_{\rm C}\gg1$.
After calibration the thermometer is ready to be operated at $\beta
E_{\rm C}\approx1$. Therefore, the thermometers might be used in the
m\,K--region and their field of application is rather narrow. Owing to
quantum fluctuations the effect will be disturbed for very low
temperatures if SET--conditions hold ($R\gg h/e^2$). In case of a single
junction a finite environment resistance will cause changes.

One possible kind of usage is as on--chip thermometer of other
SET experiments. The advantage of it is the accurate determination of
the junction temperature on the chip. Due to $\beta E_{\rm C}\approx1$
the technical restrictions of the thermometer junctions are less rigid
than those of the other SET junctions on chip, where $\beta E_{\rm
C}\gg1$ must be fulfilled. The disadvantage is the above mentioned
calibration that requires even lower temperatures than those of the
following experiment.

The presented theoretical calculations allow more detailed treatment of
the thermal behavior of SET junctions. The search for an analytical
approximation at $\beta E_{\rm C}\approx1$ might be successful. An
interesting point of the thermal behavior of small tunnel junctions
that may become important within a more detailed treatment is the
electron heating by the current~\cite{kuz3}. However, in the
standard framework applied here this effect is neglected. Eventually it
must be taken into account for lower temperatures ($\beta E_{\rm C}\gg
1$). In this case the theoretical approach basing on thermodynamic
equilibrium is questionable. It should be noted that in
this case the electron temperature exceeds the substrate
temperature~\cite{kuz3}. In Sec.~\ref{single} the study of a non--ohmic
or non--infinite environment might make sense. The investigation of an
asymmetry might be of interest in case of the double junction.
Therefore, this paper gives a first theoretical approach only.

For useful hints we are indebted to L. S. Kuzmin and S. Vyshenski. This
work was supported by the Deutsche Forschungsgemeinschaft.

\begin{figure}
\caption{Differential conductivity of a single junction within a
high--impedance environment for varying temperature showing the change
of the differential conductivity dip at $V=0$. The depth of the dip is
estimated. The curves belong to $\beta E_{\rm C} =
0.01$, $0.02$, $0.05$, $0.1$, $0.2$, $0.5$, $1.0$, $2.0$, $5.0$, $10$,
$20$, $50$, and $100$.}
\label{fig1}
\end{figure}

\begin{figure}
\caption{Minimal differential conductivity $G(V=0)$ of a
single junction in a high--impedance environment. The solid line links
the values taken from Fig.~1, the dotted line corresponds to a fit
formula given in the text. The long and short--dashed curves show
analytical approximations for $\beta E_{\rm C}\ll 1$ and $\beta E_{\rm
C}\gg 1$.}
\label{fig2}
\end{figure}

\begin{figure}
\caption{Differential conductivity of a symmetric double junction in
dependence on the applied voltage for temperatures $\beta E_{\rm C} =
0.1$, $0.2$, $0.5$, $1.0$, $2.0$, $5.0$, $10$, $20$, $50$, $100$. The
peaks at low temperatures arise from the Coulomb staircase.}
\label{fig3}
\end{figure}

\begin{figure}
\caption{Minimal differential conductivity $G(V=0)$ of a symmetric
double junction. Again, the solid line belongs to the data of Fig.~3
and the dotted line to a fit. The long and short--dashed curves show
analytical approximations for $\beta E_{\rm C}\ll 1$ and $\beta E_{\rm
C}\gg 1$.}
\label{fig4}
\end{figure}

\begin{figure}
\caption{Actual occupation probability $\sigma(n)$ of $n$ charges on the
central electrode of a symmetric double junction and a corresponding
gaussian probability distribution for varying temperatures $(a)$ $\beta
E_{\rm C} = 0.1$, $(b)$ $1.0$, $(c)$ $10$. For $\beta\to0$ the Gaussian
becomes exact.}
\label{fig5}
\end{figure}

\begin{figure}
\caption{Influence of the incoherent co--tunneling on the differential
conductance $G(V=0)$ for different values of $R/R_{\rm Q}$. As in
Fig.~4, the dotted line corresponds to the Fit to guide the eye. The
other curves show the total differential conductance in
high--temperature approximation at $V=0$ for the ratios $R/R_{\rm Q} =
1$, $10$, $100$, $1000$ from up to down.}
\label{fig6}
\end{figure}

\end{document}